\documentclass[aps,prd,preprint,longbibliography,onecolumn,showpacs,amsmath,amssymb,superscriptaddress,nofootinbib,floatfix,preprintnumbers,secnumarabic]{revtex4-2}

\usepackage{dcolumn}
\usepackage{bm}
\usepackage[T1]{fontenc}
\usepackage{graphicx}
\usepackage{amssymb,amsmath}
\usepackage{booktabs}
\usepackage{multirow}
\usepackage[dvipsnames]{xcolor}
\def\linkcolor{cyan!70!black}

\usepackage[
colorlinks=true
,urlcolor=\linkcolor
,anchorcolor=\linkcolor
,citecolor=\linkcolor
,filecolor=\linkcolor
,linkcolor=\linkcolor
,menucolor=\linkcolor
,linktocpage=true
,pdfproducer=medialab
,pdfa=true
]{hyperref}
\usepackage{tikz}
\usepackage{mathtools}
\usepackage[capitalise]{cleveref}
\usepackage{cancel}
\usepackage{feynmp}
\usepackage{soul}
\usepackage{comment}

\usepackage[normalem]{ulem}

\usepackage{slashed}
\usepackage{epsfig,psfrag,rotating,soul}
\usepackage{rotfloat}
\usepackage[font={small},justification=Justified, singlelinecheck=false,format=plain]{caption}
\usepackage{colortbl}
\usepackage{soul}
\usepackage{xtab}

\definecolor{lime}{HTML}{A6CE39}
\DeclareRobustCommand{\orcidicon}{%
    \begin{tikzpicture}
    \draw[lime, fill=lime] (0,0) 
    circle [radius=0.16] 
    node[white] {{\fontfamily{qag}\selectfont \tiny ID}};
    \draw[white, fill=white] (-0.0625,0.095) 
    circle [radius=0.007];
    \end{tikzpicture}
    \hspace{-2mm}
}
\newcommand{\orcid}[1]{\href{https://orcid.org/#1}{\orcidicon}}

\newcommand{\be}{\begin{equation}}
\newcommand{\ee}{\end{equation}}

\newcommand{\beq}{\begin{equation}} 
\newcommand{\eeq}{\end{equation}} 
\newcommand{\ba}{\begin{array}}  
\newcommand{\ea}{\end{array}} 
\newcommand{\bea}{\begin{eqnarray}}  
\newcommand{\eea}{\end{eqnarray} }  
\newcommand{\bal}{\begin{align}}
\newcommand{\eal}{\end{align}}   
\newcommand{\bi}{\begin{itemize}}  
\newcommand{\ei}{\end{itemize}}  
\newcommand{\ben}{\begin{enumerate}}
\newcommand{\een}{\end{enumerate}}  
\newcommand{\bc}{\begin{center}}
\newcommand{\ec}{\end{center}} 
\newcommand{\bt}{\begin{table}}
\newcommand{\et}{\end{table}}  
\newcommand{\btb}{\begin{tabular}}
\newcommand{\etb}{\end{tabular}}

\allowdisplaybreaks

\let\OLDthebibliography\thebibliography
\renewcommand\thebibliography[1]{
  \OLDthebibliography{#1}
  \setlength{\parskip}{0pt}
  \setlength{\itemsep}{0pt plus 0.3ex}
}

\usepackage{fontawesome5}
\makeatletter
\newcommand{\github}[1]{%
   \href{#1}{\faGithubSquare}%
}
\makeatother

\begin{document}

\begin{titlepage}

\begin{flushright}
    IFT-UAM/CSIC-24-188
\end{flushright}

\title{Disentangling axion-like particle couplings to nucleons via a delayed signal in Super-Kamiokande from a future supernova}


\author{David Alonso-Gonz\'alez\orcid{0000-0002-7572-9184}}\email{david.alonsogonzalez@uam.es}
\affiliation{Instituto de F\' \i sica Te\'orica, IFT-UAM/CSIC, 28049 Madrid, Spain}
\affiliation{Departamento de F\' \i sica Te\'orica, Universidad Aut\'onoma de Madrid, 28049 Madrid, Spain}

\author{David Cerde\~no\orcid{0000-0002-7649-1956}}
\email{davidg.cerdeno@gmail.com}
\affiliation{Instituto de F\' \i sica Te\'orica, IFT-UAM/CSIC, 28049 Madrid, Spain}

\author{Marina~Cerme\~no\orcid{0000-0001-6881-7285}}
\email{marina.cermeno@ift.csic.es}
\affiliation{Instituto de F\' \i sica Te\'orica, IFT-UAM/CSIC, 28049 Madrid, Spain}

\author{Andres D. Perez\orcid{0000-0002-9391-6047}}\email{andresd.perez@uam.es}
\affiliation{Instituto de F\' \i sica Te\'orica, IFT-UAM/CSIC, 28049 Madrid, Spain}
\affiliation{Departamento de F\' \i sica Te\'orica, Universidad Aut\'onoma de Madrid, 28049 Madrid, Spain}

\date{\today}

\begin{abstract}
In this work, we show that, if axion-like particles (ALPs) from core-collapse supernovae (SNe) couple to protons, they would produce very characteristic signatures in neutrino water Cherenkov detectors through their scattering off free protons via $a \, p \rightarrow p \, \gamma$ interactions. Specifically, sub-MeV ALPs would generate photons with energies $\sim 30$ MeV, which could be observed by Super-Kamiokande and Hyper-Kamiokande as a delayed signal after a future detection of SN neutrinos. We apply this to a hypothetical neighbouring SN (at a maximum distance of 100~kpc) and demonstrate that the region in the parameter space with ALP masses between $10^{-4}$~MeV and $1$~MeV and ALP-proton couplings in the range $3 \times 10^{-6}-4 \times 10^{-5}$ could be probed. We argue that this new signature, combined with the one expected at $\sim 7$~MeV from oxygen de-excitation, would allow us to disentangle ALP-neutron and ALP-proton couplings.
\end{abstract}

\preprint{IFT-UAM/CSIC-24-188}


\maketitle
\thispagestyle{empty}

\end{titlepage}

\section{Introduction}\label{sec:intro}

Core collapse supernovae (SNe) are the spectacular last stage in the life of giant stars. The extreme temperatures and densities reached in their interiors make these objects an exceptional natural laboratory for the study of new physics. Galactic SNe are particularly interesting because their proximity allows for a unique opportunity to directly measure signals from SN neutrinos and other exotics such as axion-like particles (ALPs).

ALPs are hypothetical pseudoscalar particles predicted in several extensions of the Standard Model~\cite{Svrcek:2006yi, Gelmini:1980re, Davidson:1981zd, Wilczek:1982rv, Cicoli:2012sz, Halverson:2019kna}. They are pseudo-Nambu Goldstone bosons, as the QCD axion \cite{Peccei:1977hh1, Peccei:1977hh2, Weinberg:1977ma, Wilczek:1977pj}, but contrary to these, their mass and couplings to SM particles are, in general, free parameters. ALPs with masses below $\mathcal{O}(100)$~MeV can be produced copiously in SN explosions. This has been explored for models where ALPs couple to nucleons in Refs.~\cite{RAFFELT19901, Raffelt:1993ix, Raffelt:1996wa, Fischer:2016cyd, Carenza:2019pxu, Carenza:2020cis, Fischer:2021jfm, Lella:2022uwi, Lucente:2022vuo, Lella:2023bfb, Carenza:2023lci, Chakraborty:2024tyx, Lella:2024dmx}. If ALPs (due to their very small coupling to SM particles) escape the SN unimpeded, they could take away with them a significant fraction of the proto-neutron star (proto-NS) energy, shortening the duration of the expected neutrino burst so significantly that it would be inconsistent with the neutrino observations from the renowned SN 1987A~\cite{burrows1989}. This has been used to constrain the ALP-nucleon coupling~\cite{PhysRevLett.60.1797, PhysRevLett.60.1793, PhysRevD.42.3297, RAFFELT19901, Fischer:2021jfm,Lella:2022uwi, Lella:2023bfb}, excluding values between $10^{-9}$ and $10^{-6}$ for ALP masses up to $\sim 100$~MeV~\cite{Carenza:2023lci}. These are usually referred to as cooling bounds. Likewise, the absence of additional events in Kamiokande-II from oxygen excitation induced by ALPs from the SN 1987A has provided complementary limits on their parameter space. These observations have excluded ALP-nucleon couplings above cooling bounds for ALP masses below $\sim 10^{-3}$~MeV~\cite{Carenza:2023wsm, Lella:2023bfb, Carenza:2023lci}.

In a recent work~\cite{Alonso-Gonzalez:2024ems}, we identified a new potential signal from ALPs that couple to protons in neutrino water Cherenkov detectors from their scattering off free protons, $a\, p\to p\, \gamma$. This process leaves a characteristic feature at energies greater than $20$~MeV, where the backgrounds are very small. We showed that ALPs with masses above $1$~MeV would be emitted semi-relativistically in SN explosions, generating a diffuse galactic SN ALP flux. Using data from Super-Kamiokande (SK)~\cite{Super-Kamiokande:2021jaq}, we derived bounds in the ALP mass range of $1-80$~MeV, excluding ALP-proton couplings between $2\times10^{-4}$ and $6\times10^{-6}$.

In this article, we return to this same process, $a\, p\to p\, \gamma$, but we now apply it to the study of ALPs with masses below $1$ MeV, focusing on a region of the parameter space above cooling bounds that has not been excluded yet. We demonstrate that ALPs from a future nearby SN can produce photons with a peak in their energy spectrum at $\sim 30$ MeV, which could be observed by SK and Hyper-Kamiokande (HK)~\cite{Hyper-Kamiokande:2018ofw}. Due to the massive nature of the ALP, this signature would be delayed relative to the observation of SN neutrinos. Furthermore, since this new signature is only sensitive to the ALP-proton coupling, and the oxygen de-excitation one~\cite{Engel:1990zd, Carenza:2023wsm} depends on both the ALP-proton and ALP-neutron couplings, we claim that the observation of both features could be used to disentangle both couplings in models with ALP-nucleon interactions.

This article is organized as follows. In Section~\ref{sec:ALP_SN}, we introduce the Lagrangian that describes ALP-nucleon interactions and the expression of the ALP flux at Earth from a single nearby SN. In Section~\ref{sec:detection}, we describe the calculation of the number of photon events expected in neutrino water Cherenkov detectors produced by ALPs. We evaluate the detection prospects of the SK and HK detectors and propose a strategy to disentangle the ALP couplings to protons and neutrons in the event of an observable SN. Finally, in Section~\ref{sec:conclusions}, we present our conclusions.

\section{ALPs from a single supernova}\label{sec:ALP_SN}

Seconds after SN core collapse, the resulting proto-NS is compressed to extreme densities, $\rho \sim 3\times 10^{14} \, \rm g/cm^3$, and heated up to temperatures of the order of $T \sim 30$ MeV~\cite{fischer2012, Fischer:2021jfm}. In this environment, if ALPs couple to nucleons through the interaction Lagrangian~\cite{PhysRevD.40.652,DiLuzio:2020wdo,Chang:1993gm, Lella:2024dmx}
\begin{align}
    \mathcal{L}_{int} = g_a \, \frac{\partial_{\mu} a}{2 \, m_N} \, \bigg[ 
    & C_{ap} \, \bar{p} \, \gamma^{\mu} \, \gamma_5 \, p 
    + C_{an} \, \bar{n} \, \gamma^{\mu} \, \gamma_5 \, n \notag \\
    & + \frac{C_{a \pi N}}{f_{\pi}} \, \left( i \, \pi^+ \, \bar{p} \, \gamma^{\mu} \, n - i \, \pi^- \, \bar{n} \, \gamma^{\mu} \, p \right) \notag \\
    & + C_{a N \Delta} \, \left( \bar{p} \, \Delta^+_{\mu} + \bar{\Delta^+_{\mu}} \, p 
    + \bar{n} \, \Delta^0_{\mu} + \bar{\Delta^0_{\mu}} \, n \right) 
    \bigg],
    \label{eq:Laxion}
\end{align}
they can be produced via nucleon-nucleon Bremmstrahlung, $N N \to N N a$, and pion-ALP conversions, $\pi\, N \rightarrow N\, a$, in the proto-NS core~\cite{RAFFELT19901, Raffelt:1993ix, Raffelt:1996wa, Giannotti:2005tn, Fischer:2016cyd, Carenza:2019pxu,PhysRevLett.60.1793,PhysRevLett.60.1797,burrows1989,PhysRevD.42.3297,Carenza:2019pxu,Carenza:2020cis,Fischer:2021jfm,Choi:2021ign, Lella:2022uwi, Lucente:2022vuo, Lella:2023bfb, Carenza:2023lci, Chakraborty:2024tyx, Lella:2024dmx, Alonso-Gonzalez:2024ems}.  In Eq.~\eqref{eq:Laxion}, $p$ and $n$ denote the proton and the neutron, $\pi$ the pion, and $\Delta$ the $\Delta$ baryon. The coupling $g_a$ is a dimensionless constant that is related to the ALP scale $f_a$ as $g_a = m_N / f_a$, with $m_N =938$ MeV the nucleon mass. $f_{\pi}=92.4$ MeV is the pion decay constant, and $C_{aN}$ are model-dependent $O(1)$ ALP-nucleon coupling constants with $N=p, n$. The ALP-pion-nucleon and the ALP-nucleon-$\Delta$ baryon couplings can be written as $C_{a \pi N} = (C_{ap} - C_{an}) / \sqrt{2} g_A$~\cite{Choi:2021ign}, and $C_{a N \Delta} = - \sqrt{3}/2 (C_{ap} - C_{an})$, with $g_A \simeq 1.28$~\cite{PDG:2022} the axial coupling.

As shown in Refs.~\cite{Lella:2023bfb, Alonso-Gonzalez:2024ems}, for couplings above $g_{aN} \sim 10^{-8}$, ALPs are diffusively trapped inside the proto-NS core, and absorption effects via $N\, N\, a \rightarrow N\, N$ and $N\, a \rightarrow N\, \pi$ processes become significant. In this regime, the ALP flux at Earth from a SN located at a distance $d_{SN}$, integrated over the time interval after core collapse during which the majority of ALP production occurs, can be expressed as  
\begin{equation}
    \frac{d\Phi_a}{dE_a^{\mathrm{Earth}}} = \frac{1}{4 \pi d_{SN}^2} \int_{t_{\rm min}}^{t_{\rm max}} dt \int_0^{\infty} \alpha(r)^{-1} 4 \pi r^2 dr \, \left\langle e^{-\tau\left(E_a^*, t,  r\right)} \right\rangle \frac{d^2 n_a}{dE_a^{\rm loc} dt} \left( r, t, \alpha(r)^{-1} E_a^{\mathrm{Earth}} \right).
    \label{eq:flux}    
\end{equation}  
Here, $E_a^{\mathrm{Earth}} = \alpha(r) E_a^{\rm loc}$ is the observed energy at Earth, redshifted relative to the local energy, $E_a^{\rm loc}$, through the lapse function $\alpha(r) \leq 1$, which depends on the radial position with respect to the center of the core, $r$, and encodes the effects of the proto-NS gravitational potential $\Phi(r)$, see Ref.~\cite{Alonso-Gonzalez:2024ems} for more details. The total ALP production rate $ \frac{d^2 n_a}{dE_a^{\rm loc} dt}$ accounts for nucleon-nucleon bremsstrahlung and pion-ALP conversion processes, whose production rates have been derived in Refs.~\cite{RAFFELT19901, Raffelt:1993ix, Raffelt:1996wa, Giannotti:2005tn, Carenza:2019pxu,Carenza:2020cis, Choi:2021ign} (Ref.~\cite{Carenza:2023wsm} provides a very useful compilation of these expressions).

Absorption effects are incorporated through the exponential suppression term $\left\langle e^{-\tau\left(E_a^*, t, r\right)} \right\rangle$, where $\tau\left(E_a^*, t, r\right)$ is the optical depth evaluated at $E_a^* = E_a^{\rm loc} \alpha(r)/\alpha(\sqrt{r^2+s^2+2rs\mu})$,
which is the ALP energy after accounting for the gravitational redshift between the ALP production and absorption points. Further details on the computation of this quantity can be found in Refs.~\cite{Carenza:2023wsm, Carenza:2023lci}.

Note that the radial and time dependences in $\frac{d^2 n_a}{dE_a^{\rm loc} dt}\left(r, t, E_a^{\rm loc} \right)$, $\tau\left(E_a^*, t, r\right)$, and $\alpha(r)$, arise from their dependences on the proto-NS temperature, density, and other related quantities.

\section{Signal in neutrino water Cherenkov detectors}\label{sec:detection}

\begin{figure}[!t]
  \centering
  \includegraphics[width=.4\textwidth]{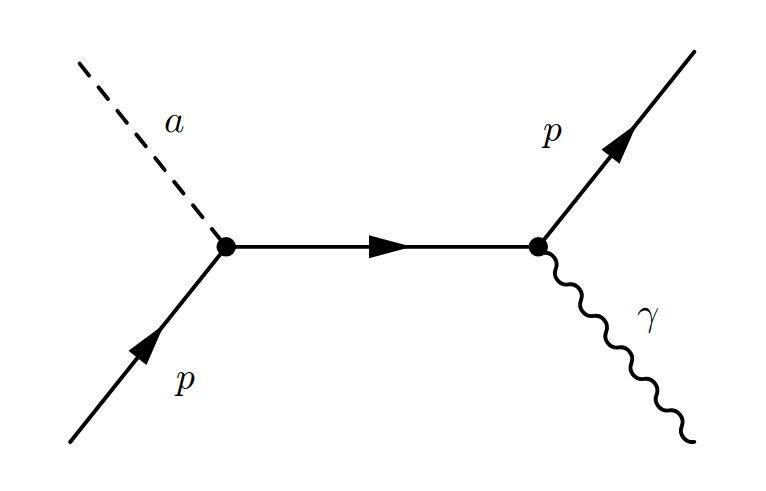}
  \includegraphics[width=.4\textwidth]{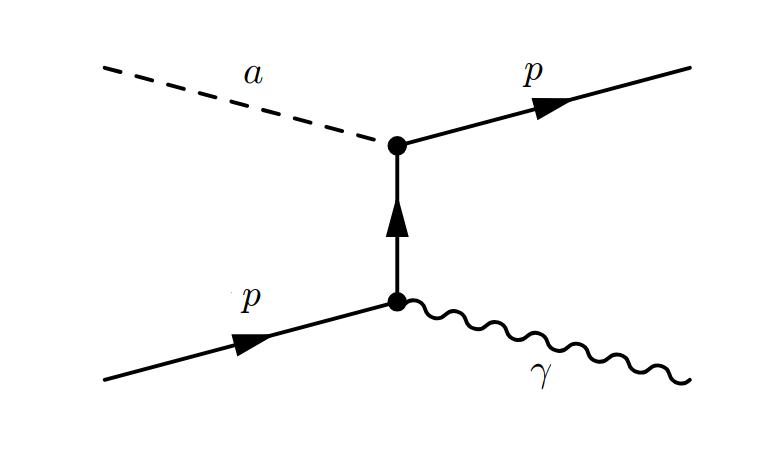}
  \caption{Photo-production via ALP-proton interaction diagrams.}
  \label{fig:feynman}
\end{figure}

The ALP flux of a single SN can be detected in neutrino water Cherenkov detectors due to its interactions with free protons via the process $a\,p\rightarrow p\,\gamma$, through the diagrams depicted in Fig.~\ref{fig:feynman}. The differential cross-section for the interaction between an ALP with energy $E_a$ scattering on a proton can be computed as
\begin{equation}
    \frac{d\sigma_{ap}}{dE_\gamma}= \frac{1}{32\pi} \int_{-1}^{1} \frac{\left|\overline{\mathcal{M}}\right|_{ap}^2}{\left(E_a^2-m_a^2 \right) m_p}\ \delta(\cos\theta-\cos\theta_0)\ d\textrm{cos}\,\theta,
   \label{eq:dxs} 
\end{equation}
where the averaged squared amplitude of the process is
\begin{align}
 |\overline{\mathcal{M}}|_{ap}^2= & \frac{C_{ap}^2 e^2 g_a^2}{E_\gamma^2 m_p \left(2 E_a m_p+m_a^2\right)^2} \; \bigg[ 4 m_p^3 (E_a-E_\gamma)^2 \left(2 E_a E_\gamma+m_a^2\right) \notag \\ & +4 m_a^2 m_p^2
		(E_a-E_\gamma) \left(E_\gamma (E_a-E_\gamma)+m_a^2\right) \notag \\  &+ m_a^4 m_p \left(2 E_a E_\gamma
		+m_a^2\right)+E_\gamma m_a^6 \bigg] \, ,
\end{align}
with $E_\gamma$ the energy of the outgoing photon and $m_p$ the proton mass. In Eq.~\eqref{eq:dxs}, $\cos\theta_0$ is the cosine of the angle between the photon and the ALP momenta, which is fixed by energy conservation, setting the minimum and maximum allowed energies for the photon produced by an ALP with energy $E_a$,
\begin{equation}
   E_\gamma^{\rm min}=\frac{m_a^2 + 2 E_a m_p}{2 (m_p + E_a) - 2 \sqrt{E_a^2 - m_a^2}},
	\,\,\,\,\,\,\,\,E_\gamma^{\rm max}=\frac{m_a^2 + 2 E_a m_p}{2 (m_p + E_a) + 2 \sqrt{E_a^2 - m_a^2}}. 
    \label{eq:elimits}
\end{equation} 
Similarly, the maximum and minimum ALP energy for a fixed photon energy, $E_a^{\rm min}(E_\gamma)$ and $E_a^{\rm max}(E_\gamma)$, can be obtained. For further details, see Ref.~\cite{Alonso-Gonzalez:2024ems}.

The differential photon spectrum produced at an experiment with $N_t$ targets can be computed as
\begin{equation}
    \frac{d N_\gamma}{dE_\gamma}=N_t \int_{E_a^{\rm min}(E_\gamma)}^{E_a^{\rm max}(E_\gamma)} dE_a^{\mathrm{Earth}}\frac{d\Phi_a}{dE_a^{\mathrm{Earth}}} \frac{d\, \sigma_{ap}}{d\, E_\gamma},
    \label{eq:photon_spectrum}
\end{equation}
where $\frac{d\Phi_a}{dE_a^{\mathrm{Earth}}}$ is the flux of ALPs at Earth generated by a single SN event, shown in Eq.~\eqref{eq:flux}. Then, the number of events in a detector can be obtained simply by integrating in the experimental energy region.
\begin{figure}[!t]
  \centering
  \includegraphics[width=1.\textwidth]{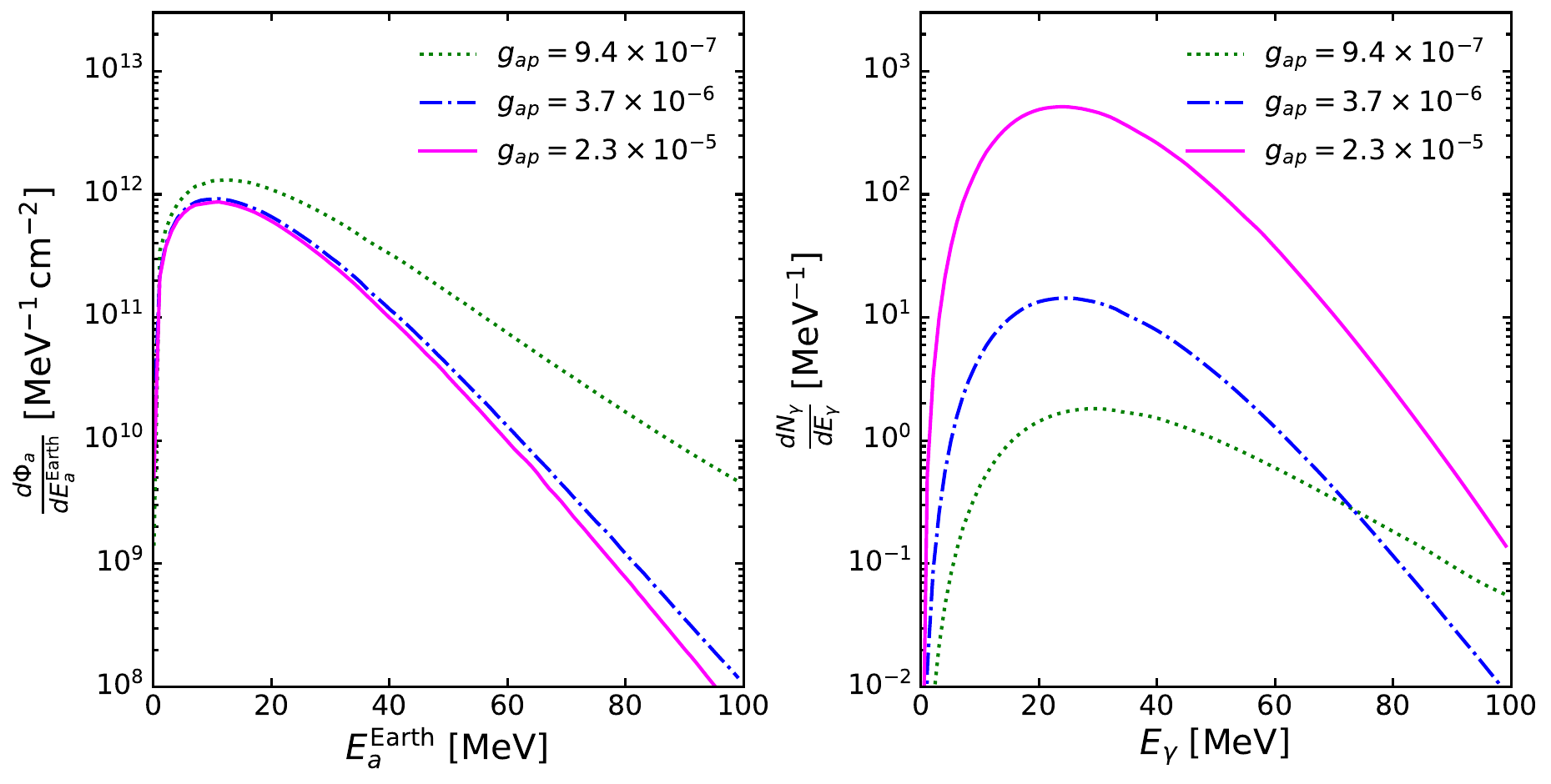}
  \caption{ALP flux reaching Earth from a SN at distance $d_{SN}=1$~kpc vs ALP-energy (left) and differential photon spectrum produced by the ALP flux in  Super-Kamiokande (right) for $m_a \leq 1$~MeV and $g_{ap}=9.4 \times 10^{-7},\, 3.7 \times 10^{-6}, \, 2.3 \times 10^{-5}$ in green dotted, blue dashed-dotted and magenta solid lines, respectively.
  }
  \label{fig:flux_events}
\end{figure}

For concreteness, throughout the rest of this article we set $C_{an}=0$, which leads to $g_{an}=0$, following the analogy with the KSVZ axion~\cite{Kim:1979if, Shifman:1979if}. This model has been previously analyzed in the context of SNe, as discussed, for instance, in Refs.~\cite{Lella:2022uwi, Lella:2023bfb, Carenza:2023lci, Alonso-Gonzalez:2024ems}.

To characterize the SN environment, we consider a $18 M_\odot$ progenitor simulated in spherical symmetry using the AGILE-BOLTZTRAN code \cite{agile1,agile2}. The temperature, density, and electron and muon fraction profiles, evaluated at 1~second post-bounce, are obtained from Ref.~\cite{Fischer:2021jfm}. The ALP flux is computed by integrating over a time interval of $1.5$~seconds, corresponding to $t_{\rm min}=0.5$~s and $t_{\rm max}=2$~s in Eq.~\eqref{eq:flux}, as these profiles remain approximately constant during this period (see, e.g., Fig.~7 in Ref.~\cite{fischer2012}). Note that this approach yields a conservative estimate of the ALP signal expected in neutrino detectors.

On the left panel of Fig.~\ref{fig:flux_events}, we present the ALP flux at Earth from a potential future SN located at a distance $d_{SN}=1$~kpc, for three different ALP-proton couplings, and assuming $m_a \leq 1$~MeV. For these couplings, which fall within the trapping regime, the flux peaks at energies around $10$~MeV. For smaller couplings, i.e., $g_{ap} \lesssim 10^{-8}$, ALPs escape the proto-NS freely, leading to a peak at higher energies, approximately around $50$~MeV \cite{Lella:2023bfb, Carenza:2023wsm, Alonso-Gonzalez:2024ems}. Conversely, for larger couplings, ALPs become diffusively trapped, shifting the flux peak to lower energies. Absorption effects become evident when comparing the fluxes for different couplings: stronger couplings yield lower values for the flux, since a higher amount of ALPs are re-absorbed. It is also worth noting that for $m_a \leq 1$~MeV, ALPs can be considered effectively massless, and the flux remains independent of $m_a$.

On the right panel of Fig.~\ref{fig:flux_events}, we show the differential photon spectrum from $a\, p \rightarrow p \, \gamma$ expected at SK for the same couplings and range of masses as on the left panel. This is obtained after convoluting the ALP flux with the differential cross section, as per Eq.~\eqref{eq:photon_spectrum}, and results in a shift and broadening of the peak, which reaches its highest value at $\sim 30$~MeV. In contrast to the flux shown on the left panel, the photon event spectrum increases with the coupling, due to the cross-section dependence on the coupling constant, which scales as $g_{ap}^2$.

\subsection{Prospects of detection in Super-Kamiokande}

\begin{figure}[!t]
  \centering
  \includegraphics[width=1.\textwidth]{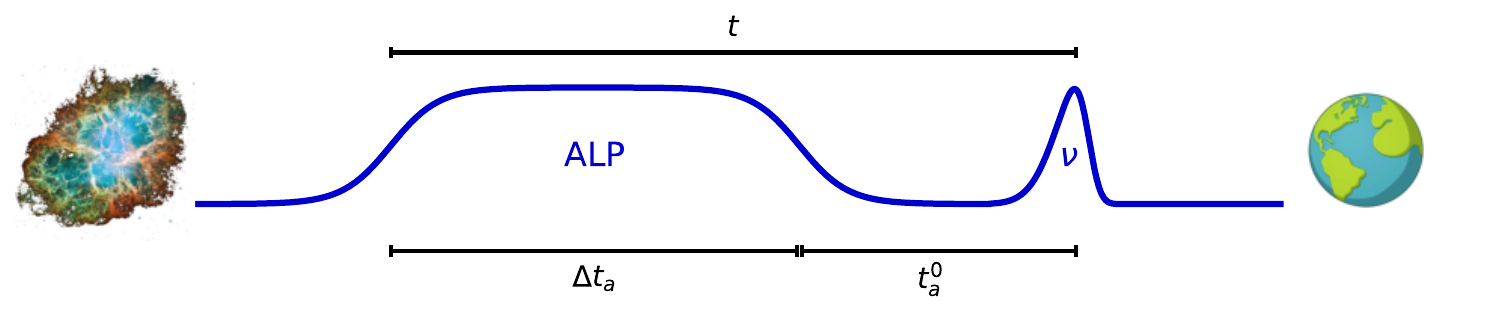}
  \caption{Schematic time behaviour of the neutrino and ALP package reaching Earth after being produced in a SN event.
  }
  \label{fig:schemetime}
\end{figure}

In a future SN, ALPs would be produced at the same time as neutrinos. However, since ALPs are massive, they travel more slowly, leading to a time delay between their arrival and the first neutrino event. This is schematically represented in \cref{fig:schemetime}. We define $t_a$ as the time difference between the arrival of an ALP with energy $E_a$ and the neutrinos, which can be estimated as \cite{Raffelt:1996wa, Lella:2023bfb}
\begin{equation}
    t_{a} \, \simeq \, \frac{d_{SN} \, m_a^2}{2 E_a} \sim 2.01 \times10^6 \, \text{s} \, \bigg( \frac{d_{SN}}{1 \, \text{kpc}} \bigg) \, \bigg( \frac{m_a}{0.1 \, \text{MeV}} \bigg)^2 \, \bigg( \frac{16 \, \text{MeV}}{E_a} \bigg)^2 \, . 
    \label{eq:timeALP}
\end{equation} 
The first ALP (the most energetic one) would therefore arrive at $t_{a}^0$ and less energetic ALPs would arrive later due to the energy-dependent time of flight. This translates into a time window, $\Delta t_a$, of arrival of the ALP package with energy range $[E_a^{\rm low}, E_a^{\rm high}]$, 
\begin{equation}
    \Delta t_{a} \, \simeq \, t_a (E_a^{\rm low}, m_a, d_{SN}) \, - \, t_a (E_a^{\rm high}, m_a, d_{SN})  \, .
    \label{eq:timewindow}
\end{equation}
Thus, $t=t_a^0+\Delta t_a$ is the total time that water Cherenkov experiments should be taking data in order to measure both the neutrino signal and the whole ALP package.

To compare the expected number of signal and background events during the time window $\Delta t_a$, we have defined the signal significance as 
\begin{equation}
   Z(\Delta t_a) \, = \, \frac{N_{\gamma}(\Delta t_a)}{\sqrt{\bar{n}_{bkg} \, \Delta t_a}}\ ,
    \label{eq:significance}
\end{equation} 
where $\bar{n}_{bkg}$ is the number of background events per unit of time in SK, and $N_{\gamma}(\Delta t_a)$ and $\bar{n}_{bkg} \, \Delta t_a$ are the number of expected photon produced by ALPs and background events during the ALP time window, respectively. To account for a 95$\%$ C.L., we set $Z(\Delta t_a) \geq 2$, with the condition that at least 2 events have to be measured. Then, the parameter space that SK is able to probe satisfies
\begin{equation}
   N_{\gamma}(\Delta t_a) \, \geq \text{max} \left[2, \, 2\sqrt{\bar{n}_{bkg} \, \Delta t_a}\right].
    \label{eq:condition}
\end{equation}

Notice that $\Delta t_a$ depends on the energy range of the incoming ALPs (see Eq.~\ref{eq:timewindow}). This, in turn, determines the energy range of the resulting photons and the range of experimental reconstructed energy where the search is performed for each signal.

For the $a\, p \rightarrow p \, \gamma$ process, the photon spectrum peaks at $\sim 20-30$ MeV (see Fig.~\ref{fig:flux_events}). Thus, we have defined the region of interest in reconstructed energy $E_{rec}=[16, 78]$~MeV. This region was optimised to search for the diffuse SN background in Ref.~\cite{Super-Kamiokande:2021jaq} by identifying positrons from inverse beta decay interactions. The background rate is $\bar{n}_{bkg}=9.38\times 10^{-7}$ s$^{-1}$, dominated by cosmic ray muon spallation, electrons produced by the decays of invisible, or low energy, muons and pions, and atmospheric neutrinos. For this process, \cref{eq:elimits} leads to $E_{\gamma} \sim E_a$, and therefore we consider $E_a^{\rm low}=16$~MeV and $E_a^{\rm low}=78$~MeV to estimate $\Delta t_a$, which results in
\begin{equation}
    \Delta t_{a}^{a\,p \rightarrow p\, \gamma} \, \simeq  \, 1.93 \times10^6 \, \text{s} \, \bigg( \frac{d_{SN}}{1 \, \text{kpc}} \bigg) \, \bigg( \frac{m_a}{0.1 \, \text{MeV}} \bigg)^2. 
    \label{eq:timewindowAP}
\end{equation}

\begin{figure}[!t]
  \centering
  \includegraphics[width=\textwidth]{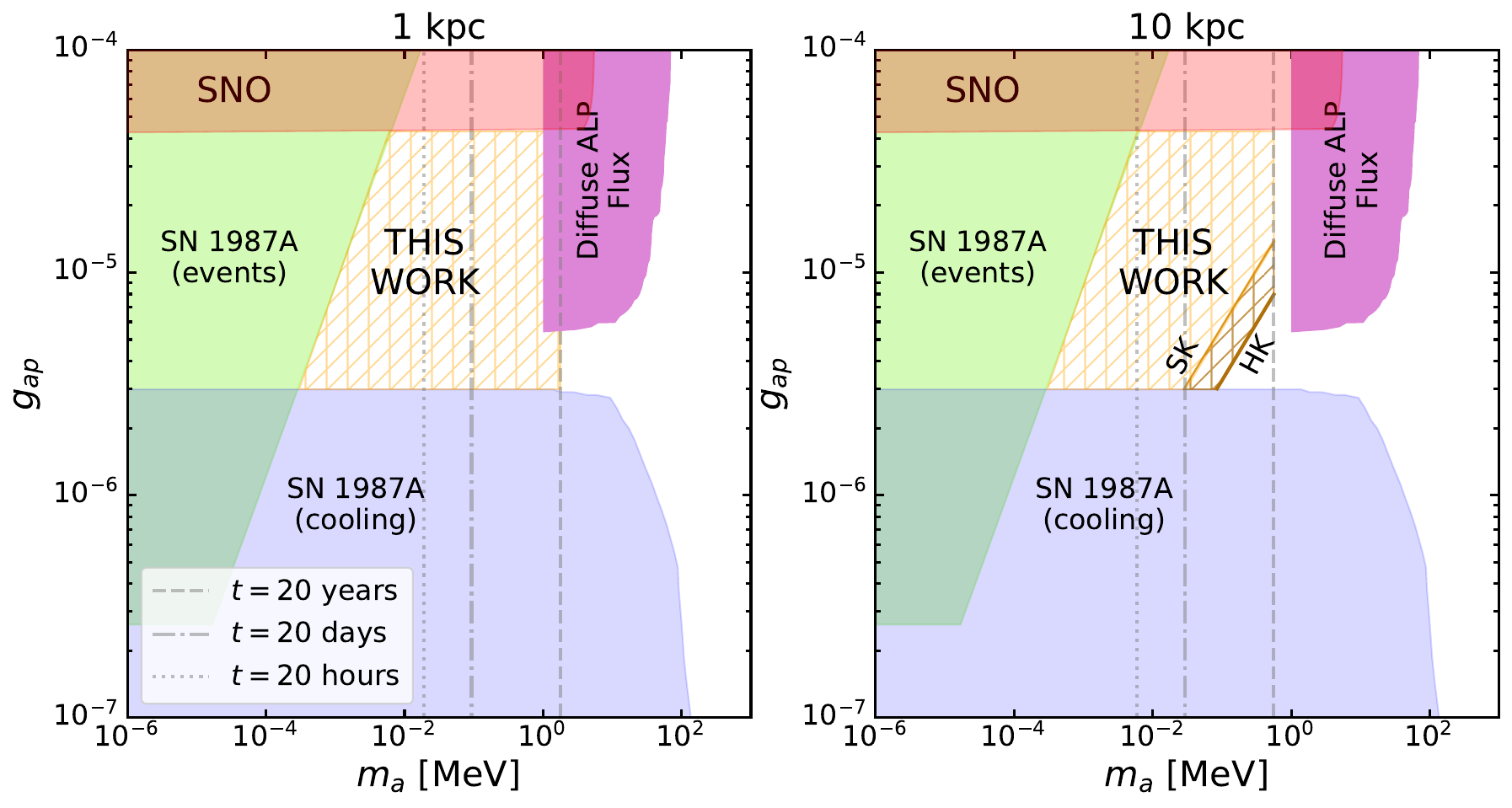}
  \caption{ALP parameter space that would be probed by SK (hatched orange) and HK (hatched brown) at 95$\%$ C.L. through $a\,p\rightarrow p\,\gamma$ from ALPs of a future SN at $1$~kpc (left) and $10$~kpc (right). Complementary bounds from solar axion events predicted in SNO~\cite{Bhusal:2020bvx} (red region), from expected events in Kamiokande-II due to oxygen de-excitation caused by ALPs from the SN 1987A~\cite{Lella:2023bfb} (green region), from the SN 1987A cooling~\cite{Lella:2023bfb} (blue region), and from the diffuse galactic SN ALP flux scattering on free protons in SK~\cite{Alonso-Gonzalez:2024ems}. Vertical grey lines indicate the ALP masses for which all emitted ALPs would take 20 years (dashed), 20 days (dashed-dotted), and 20 hours (dotted) to to reach Earth relative to the detection of the neutrino events.
  }
  \label{fig:results}
\end{figure}

In Fig.~\ref{fig:results}, the hatched orange region represents the area in the ($g_{ap}, m_a$) parameter space that can be probed by SK by the observation of photons from $a\,p\rightarrow p\,\gamma$, in the event of a SN at a distance of $d_{SN}=1$~kpc (left plot) or $d_{SN}=10$~kpc (right). For comparison, we also present current exclusion bounds from other sources.
Besides, we show as vertical grey lines the ALP mass for which all ALPs would take $t=20$ years (dashed), 20 days (dashed-dotted), and 20 hours (dotted) to arrive at Earth with respect to the detection of the first neutrino events. To be conservative, we do not analyse parameter points with $t>20$ years (the time scale that SK has been operational for). However, the analysis can easily be extended to $t > 20$ years for experiments with longer data-taking periods or to account for scenarios where only an initial fraction of the ALP package could be measured before the experiment is decommissioned.

For $d_{SN}=10$~kpc, the probed region is limited by the diagonal line between ($g_{ap}, m_a$)=($3\times10^{-6}$, $5\times10^{-2}$) and ($1.5\times10^{-5}$, $5\times10^{-1}$), which is determined by the condition $Z(\Delta t_a) =2$. If one considers HK and assumes that it will have the same characteristics and data reduction techniques as SK, this limit extends to lower couplings by rescaling signal and background due to an increase in the target size, from $22.5$~kton to $187$~kton detector mass, and is shown in hatched brown.

To summarize, the detection of a nearby SN will allow to probe the parameter space with ALP masses between $10^{-4}$~MeV and $1$~MeV for ALP-proton couplings in the range $ 3 \times 10^{-6}-4 \times 10^{-5}$ in both SK and HK, covering the available parameter region between constraints from the SN 1987A, SNO solar axion searches, and the diffuse galactic SN ALP flux.

One may wonder whether the delayed signal of ALPs from the SN 1987A via $a\,p \rightarrow p\, \gamma$ could have been observed by SK. However, we have checked that this is not the case. In the scenario that maximises the ALP signal, corresponding to $g_{ap} \simeq 4\times 10^{-5}$ and $m_a \simeq 0.45$~MeV, the first (last) ALP would have reached Earth $\sim 2.8 \; (64)$~years after the SN 1987A neutrino events. Taking into account the operation period of the first four SK phases, $9$ to $31$ years after the SN 1987A neutrinos, only part of the ALP package could have been observed, corresponding to ALPs with energies in the range $23.5-43.6$~MeV. Although the peak of the ALP spectrum is contained in this range, the statistical significance of the signal is only $Z\simeq0.5$, even before considering the $6$-year pause between SK data-taking operations. Thus, no constraints can be derived.

\subsection{Strategy to discriminate ALP-proton and ALP-neutron couplings}

ALPs with interactions described by the Lagrangian in Eq.~\eqref{eq:Laxion} can also induce oxygen nuclei excitations in neutrino water Cherenkov detectors, which can be searched for through the photons emitted to relax the system. In this case, the observed spectrum would be a collection of photons mostly distributed in the energy range $[5,10]$~MeV, corresponding to the energy transitions between the most probable excited oxygen states and its ground state. This was used to derive bounds on axion properties by contrasting Kamiokande~II data with the expectation from the SN 1987A ~\cite{Engel:1990zd}. The application to ALPs has been recently reviewed in Ref.~\cite{Carenza:2023wsm}, with bounds that can be found in Ref.~\cite{Lella:2023bfb}, which improves the computation of the ALP-oxygen cross section through the refinement of nuclear models.

It is important to remark that, for the same ALP-nucleon coupling, the value of the differential photon spectrum from oxygen de-excitation at the peak ($\sim 7$~MeV) would exceed that from the $a\,p \rightarrow p\, \gamma$ spectrum by a factor $\mathcal{O}(10^{4})$. However, a fundamental feature of our photon spectrum is that the peak will appear at higher energies ($E_\gamma\sim 30$~MeV for $m_a \leq 1$~MeV), as shown on the right panel of Fig.~\ref{fig:flux_events}. At these energies, we expect photons coming from the oxygen de-excitation to be negligible compared to our signal. This distinction arises from the fundamentally different nature of the two detection methods. In our scattering process, the kinematics —and particularly the energy of the emitted photons— are strongly influenced by the ALP energy, see Eq.~\eqref{eq:elimits}. This leads to $E_{\gamma} \sim E_a^{\rm Earth}$. Specifically, for $E_a^{\rm Earth} = 1$~MeV ($80$~MeV) one gets $ E_\gamma^{\rm max} - E_\gamma^{\rm min} \sim 10^{-3}$~MeV ($12$~MeV). In contrast, photons from the oxygen channel originate directly from nuclear de-excitation, with their energies determined by the intrinsic nuclear structure of oxygen. Specifically, the energy levels of the oxygen span the interval $[9.55, 28]$~MeV \cite{Lella:2023bfb,Carenza:2023wsm} and, since the decay probability of the oxygen is much higher for transitions between consecutive levels, this results in a de-excitation photon spectrum with energies lying mostly in the range $E_\gamma<15$~MeV, as can be seen in Fig. 8 from Ref.~\cite{Carenza:2023wsm} and Fig. 4.4 from Ref.~\cite{Carenza:2018vcb}. Looking at the latter figure, we can deduce that the value of the differential photon spectrum for oxygen de-excitation at energies around $\sim 30$~MeV is at least $2$ orders of magnitude smaller than the peak of our signal. On top of this, the differential background rate in SK for energies $\sim 7$~MeV is $4$ orders of magnitude higher than the one in the energy region where most of our events take place, see Refs.~\cite{Super-Kamiokande:2016yck, Super-Kamiokande:2021jaq}.

\begin{figure}[!t]
  \centering
  \includegraphics[width=0.45\textwidth]{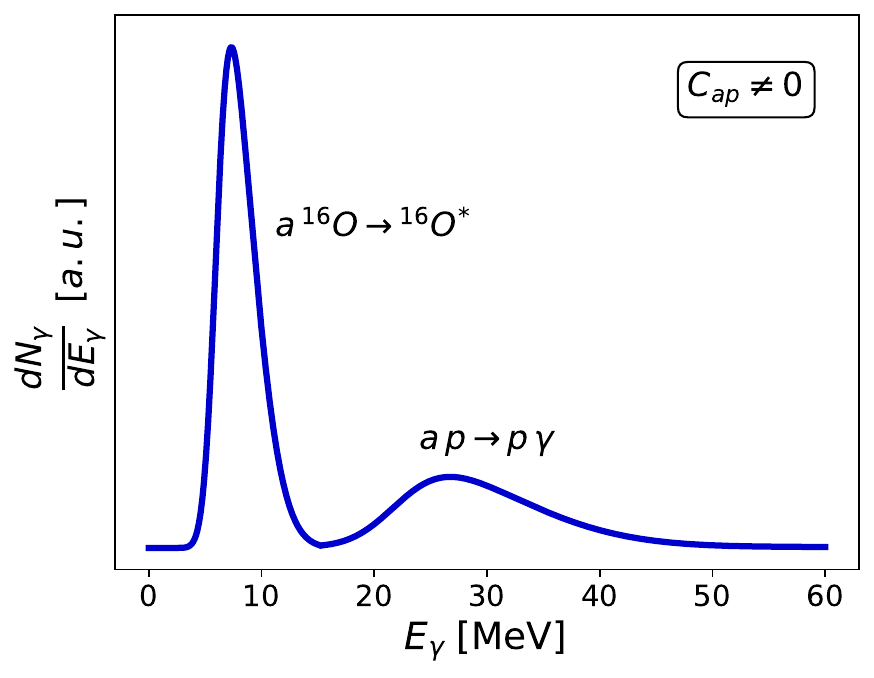}
  \includegraphics[width=0.45\textwidth]{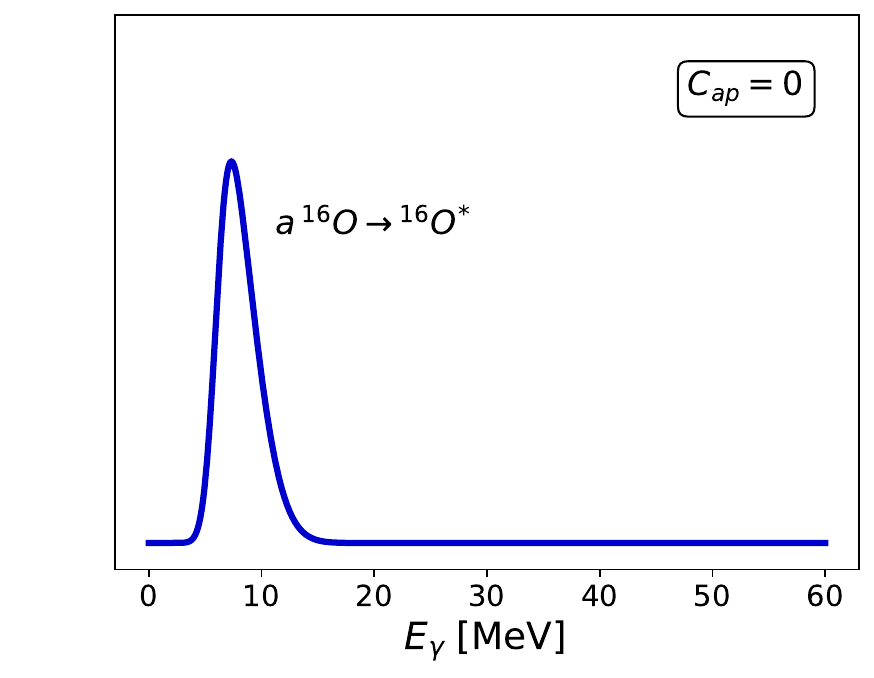}
  \caption{Schematic photon spectrum behaviour produced by ALPs from SNe. For $C_{ap} \neq 0$ (left panel) we expect a peak due to ALP-induced oxygen nuclei excitation events for low energies, and a secondary signal for higher energies corresponding to $a\, p \rightarrow p \, \gamma$ events. For $C_{ap} = 0$ (right panel) the $a\, p \rightarrow p \, \gamma$ signal is negligible.}
  \label{fig:scheme}
\end{figure}

Another interesting feature of our search channel is its exclusive sensitivity to the ALP coupling with protons, in contrast to the oxygen channel, which is determined by couplings with both protons and neutrons. This distinction could serve as a tool for disentangling the values of the two types of couplings in the event of a detection. For instance, if $g_{ap} = 0$ and $g_{an} \neq 0$, only the peak from the oxygen detection channel would be present. Conversely, if $g_{ap}$ is nonzero, both peaks would appear in the observed spectrum. In this case, the relative heights of the peaks could provide a means to determine the values of both couplings. A schematic representation of this behaviour can be seen in Fig.~\ref{fig:scheme}.

\begin{figure}[!t]
  \centering
  \includegraphics[width=.55\textwidth]{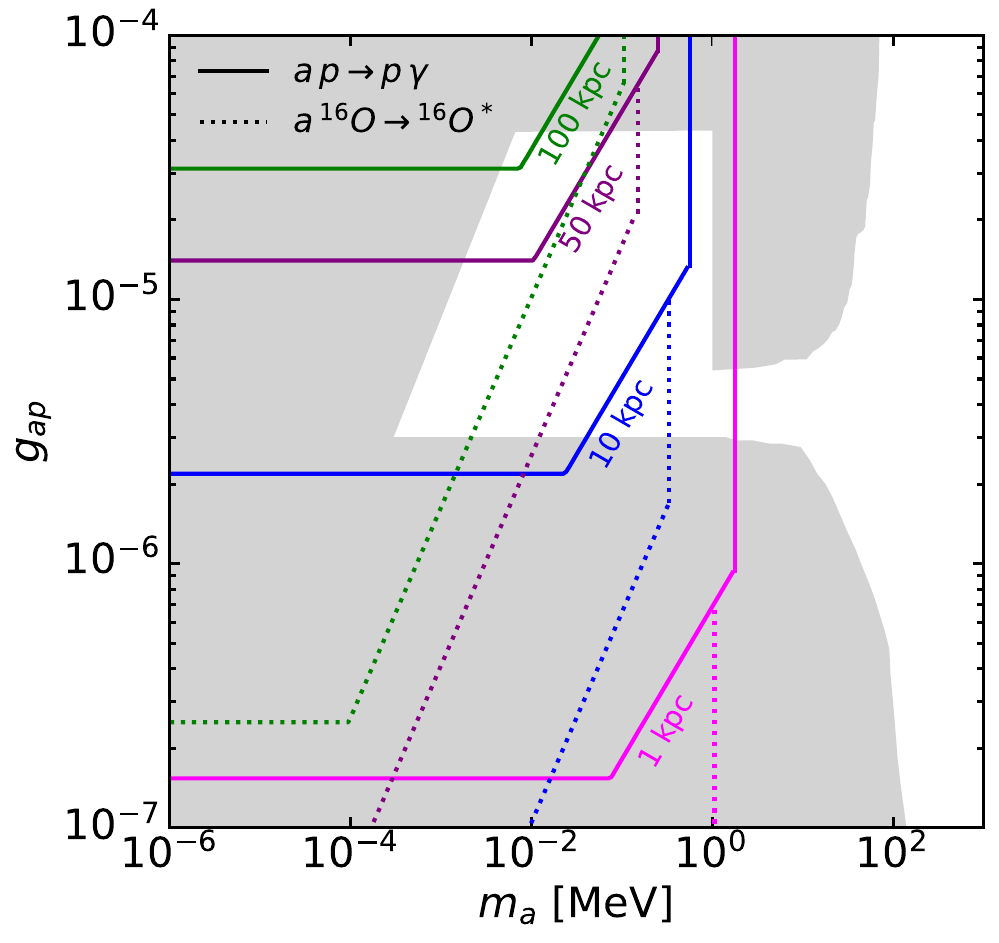}
  \caption{ALP parameter space that would be probed by Super-Kamiokande at 95$\%$ C.L. (see Eq.~\ref{eq:condition}) through the $a\,p\rightarrow p\,\gamma$ channel (solid) and our estimation for the $a\, {}^{16}O\rightarrow{}^{16}O^*$ channel (dotted) from ALPs of a future SN located at $d_{SN}=1, \, 10, \, 50, \, 100$~kpc, closed by magenta, blue, purple and green lines, respectively. Current bounds are shown in grey.}
  \label{fig:distances}
\end{figure}

To show the regions where the ALP-nucleon couplings could be discriminated, in Fig.~\ref{fig:distances} we present the ALP parameter space that SK would probe at 95$\%$ C.L. for a future SN located at $d_{SN}=1, \, 10, \, 50, \, 100$~kpc. Solid lines correspond to the $a\,p \rightarrow p\, \gamma$ channel, while dotted lines delimit the regions that could be probed by focusing only on ALP-induced oxygen nuclei excitation. In this case, as already pointed out, the resulting de-excitation photons have energies $E_{\gamma} < 15$ MeV. Thus, we consider the low-energy optimised analysis to search for solar neutrinos of Ref.~\cite{Super-Kamiokande:2016yck}, where the background is dominated by accidental coincidences and spallation events resulting in $\bar{n}_{bkg}=2.24\times 10^{-3}$~s$^{-1}$. To compute the number of signal events, we extrapolate the results presented in Ref.~\cite{Carenza:2023wsm}, which assumes an ALP energy range $[E_a^{\rm low}, E_a^{\rm high}] = [9.55, 28]$ MeV. In turn, this translates into a time window, 
\begin{equation}
    \Delta t_{a}^{a \, O \rightarrow O^*} \, \simeq \, 4.99 \times10^6 \, \text{s} \, \bigg( \frac{d_{SN}}{1 \, \text{kpc}} \bigg) \, \bigg( \frac{m_a}{0.1 \, \text{MeV}} \bigg)^2 \, , 
    \label{eq:timewindowOX}
\end{equation} 
which means that, the search strategy for oxygen-induced events not only has to be performed in a different energy range than the $a\,p \rightarrow p\, \gamma$, but also in a slightly different time window after the first neutrino event\footnote{For the nominal energy range of Ref.~\cite{Carenza:2023wsm}, oxygen-induced events would start to appear later than those from $a\,p \rightarrow p\, \gamma$. The difference depends on the ALP mass and the distance to the SN, and could be as large as several days for masses of order $m_a\sim 0.1$~MeV and a SN at $1$~kpc.}. In Fig.~\ref{fig:distances}, this effect can be seen in the vertical lines that correspond to $t=20$ years and set the ALP mass upper limits. The lower limits on the coupling are set by the condition $N_\gamma=2$. Finally, the parameter points on the diagonal lines that connect these two bounds produce $N_\gamma>2$ but $\Delta t_a$ is long enough such that the integrated background is significant and the limit is determined by $N_{\gamma}(\Delta t_a) \, \geq \, 2\sqrt{\bar{n}_{bkg} \, \Delta t_a}$. Oxygen-induced events can probe a larger parameter space than the $a\,p \rightarrow p\, \gamma$ channel, especially in coupling values, since we expect a larger number of events by $2-3$ orders of magnitude. However, we also expect a larger background by $4$ orders of magnitude, resulting in a significance $10-30$ times larger, which mainly affects the bound marked with a diagonal line. While the oxygen signal would be enough to exclude most of the parameter space, both signals are crucial to disentangle the ALP-proton and ALP-neutron coupling degeneracy in case of a positive detection.

\section{Conclusions}\label{sec:conclusions}

In this article, we have demonstrated that axion-like particles (ALPs) produced in core-collapse supernova (SN) and coupled to protons can generate a distinctive signal in neutrino water Cherenkov detectors through their interactions with free protons, $a\, p \rightarrow p\, \gamma$. The photon spectrum produced in this scattering would peak at energies $\sim 30$~MeV and could be observed by Super-Kamiokande in a region with extremely low background rate. Due to the massive nature of the ALPs, this signature would be delayed with respect to the observation of neutrinos from the SN explosion and the ALPs themselves would spread out in a package due to their different energies.

Considering a hypothetical neighbouring SN, we have computed the ALP flux at Super-Kamiokande, and extracted the regions in the ALP parameter space (mass versus coupling to protons) where the signal would be observable, as a function of the distance to the SN. For a SN located at 1 kpc from Earth, it would be possible to probe ALPs with masses up to $1$~MeV and $3\times 10^{-6} \, \leq \, g_{ap} \, \leq \, 4\times 10 ^{-5}$, covering the entire available parameter region between constraints from the SN 1987A, SNO solar ALP searches, and the diffuse galactic SN ALP flux. For more distant SNe, the sensitivity is gradually reduced towards larger couplings and lighter ALPs. We have checked that the most distant observable SN through the $a\, p \rightarrow p\, \gamma$ signal would be around 100~kpc.

We have also highlighted the potential of combining the $a\, p \rightarrow p\, \gamma$ channel (which is only sensitive to the ALP-proton coupling) with the well-known oxygen de-excitation signal, which produces another feature in the photon spectrum at lower energies, $\sim 7$ MeV, and depends on both the ALP-proton and ALP-neutron couplings. If analysed together, these complementary signals could help to disentangle the contributions from ALP-nucleon interactions.

\vspace{2.5mm}
\section*{Acknowledgments.}

We would like to thank Luis Labarga, Mario Reig and Javi Serra for useful discussions and comments. We also thank Nick Houston for details on the limits from SNO. We acknowledge support from the Spanish Agencia Estatal de Investigaci\'on through the grants PID2021-125331NB-I00 and CEX2020-001007-S, funded by MCIN/AEI/10.13039/501100011033. DGC also acknowledges support from the Spanish Ministerio de Ciencia e Innovaci\'on under grant CNS2022-135702.

\bibliographystyle{JHEP} 
\bibliography{1SN_main.bbl}

\end{document}